\documentclass[pra,letterpaper,onecolumn,showpacs,superscriptaddress]{revtex4}

\usepackage{graphicx,psfrag,amsmath,amssymb,amsfonts,bm,bbm,latexsym,color,dcolumn}

\begin{document}

\title{Equilibrium states of a test particle coupled to finite size heat baths}

\author{Qun Wei}
\affiliation{Department of Physics and Astronomy,Dartmouth College,6127 Wilder Laboratory,Hanover,NH 03755,USA}

\author{S. Taylor Smith}
\affiliation{Department of Physics and Astronomy,Dartmouth College,6127 Wilder Laboratory,Hanover,NH 03755,USA}

\author{Roberto Onofrio}
\affiliation{Dipartimento di Fisica ``Galileo Galilei",Universit\`a di Padova,Via Marzolo 8,Padova 35131,Italy}

\affiliation{Center for Statistical Mechanics and Complexity,INFM-CNR,Unit\`a di Roma 1,Roma 00185,Italy}

\affiliation{Department of Physics and Astronomy,Dartmouth College,6127 Wilder Laboratory,Hanover,NH 03755,USA}

\date{\today}

\begin{abstract}
We report on numerical simulations of the dynamics of a test particle 
coupled to competing Boltzmann heat baths of finite size. 
After discussing some features of the single bath case, we show that 
the presence of two heat baths further constrains the conditions 
necessary for the test particle to thermalize with the heat baths. 
We find that thermalization is a spectral property in which the 
oscillators of the bath with frequencies in the range of the test 
particle characteristic frequency determine its degree of thermalization. 
We also find an unexpected frequency shift of the test particle 
response with respect to the spectra of the two heat baths.
Finally, we discuss implications of our results for the study 
of high-frequency nanomechanical resonators through cold damping 
cooling techniques, and for engineering reservoirs capable of 
mitigating the back-action on a mechanical system.

\end{abstract}


\maketitle

\section{Introduction} 
Brownian motion is historically central to the molecular-level 
interpretation of statistical mechanics \cite{Einstein1,Einstein2}, 
and it is crucial for  precision measurements of macroscopic degrees 
of freedom \cite{Ornstein}.
More recently, Brownian motion has been interpreted as emerging from the
interaction of a particle, classical or quantum, with an environment
which may be schematized as a set of harmonic oscillators
\cite{CaldeiraLeggett1981,Caldeira}, representing a heat bath. As the
number of oscillators in the heat bath increases, the dynamics of the
test particle considered as an open system becomes increasingly
stochastic. Simple statistical laws for the average behavior of such
open systems are obtained by considering environments with particular
energy distributions, such as the canonical ensembles. Indeed, in the
limit of an infinite number of oscillators, and for certain additional
assumptions about their collective properties, including weak coupling
and/or high temperature limits, the equation of motion
for the test particle reduces to the familiar Langevin equation. In
this case, stationary solutions correspond to the test particle's
energy having a Boltzmann distribution in time, and therefore thermalized
with the heat bath.

In many contexts, however, in particular for systems of interest in
mesoscopic physics and nanotechnology, the size of the environment 
is small and does not justify this large number limit. 
There is also a natural infrared cut off for the frequency of 
the oscillators schematizing the environment, as the latter 
cannot support wavelengths much larger than its size. 
Furthermore, the finite amount of energy in any realistic 
enviroment demands an ultraviolet cut off in its density of states. 
Our study is aimed at understanding generic properties of thermalization
in mesoscopic systems, and we add to the analytical results direct
simulations of a test particle in contact with a finite number of
oscillators, whose frequencies are distributed in a finite bandwidth. 
This is a textbook system, but it produces a range of behaviors which 
are missed by purely analytical studies, and which lead to a spectrum of 
phenomenological implications, in particular in the physics of
high-frequency nanomechanical structures. 
Our findings may be relevant for the understanding of 
anomalies observed in the effective temperature of a resonating mode 
of a nanoresonator in interaction with a thermal bath and with the 
measurement apparatus schematized as a thermal bath at an effective 
temperature \cite{Schwab}.
The nanoresonator dynamics result from the competition between two 
effective heat baths at different temperatures, in analogy to 
the phenomenon of cold damping first discussed in \cite{Ornstein}, and 
later demonstrated for macroscopic resonators in \cite{Hirakawa}. 
Anomalies observed in the experiment described in \cite{Schwab} 
may then be attributed to deviations from the Boltzmann distribution 
for the nanoresonator.   

In a previous contribution \cite{Taylor}, conditions under which
thermalization occurs in the presence of a single finite resource heat
bath have been discussed. In this paper, after recalling the case of
thermalization in the presence of a single bath case and analyzing in 
more depth some specific issues, we continue our analysis by
considering a situation in which the test particle is intermittently interacting 
with two different reservoirs. For Langevin reservoirs, this has been 
discussed, especially in recent years, in the literature 
\cite{Kipnis,Hanggi,Kohler1,Trimper,Kohler2,Visco, Bertini}. 
Langevin reservoirs, however, are only parameterized by the temperatures of the
heat baths and the damping coefficients (or, alternatively, the relaxation times). 
In our case, the density of states of the heat baths, the number of
particles in each bath, and the masses of the constituent particles are further parameters 
which influence the thermalization of the test particle. 
When two heat baths are present, they can also differ in
temperature and density of states, opening up an even richer scenario.

The paper is organized as follows. In Section II we discuss analytical 
results on heat baths, in the two extreme situations of a heat bath 
with infinite bandwidth and of zero bandwidth. 
Both cases can be treated analytically with simple approximations, and
they both provide useful insights for the more general case of a finite bandwidth. 
In Section III we describe in some detail the numerical techniques and some 
results, complementing the ones presented  in \cite{Taylor}, about the 
dependence of thermalization upon the density of states and the
approximate scaling of the thermalization curves. 
In Section IV we present the case of the two baths and examples of
both thermalization and frustration of the test particle under 
intermittent interaction of the heat baths. 
The potential relevance of these studies to the analysis of recent experiments and
possible future demonstrations is discussed in the conclusions, with
particular regard to the possibility to engineer some baths to achieve
approximated nondemolitive measurements in both the classical and
quantum regime.

\section{Single heat bath: analytical results}

Our starting point is the classic Hamiltonian of the test particle
plus environment found in \cite{Ford1,CaldeiraLeggett1981,Caldeira,Ford2}:

\begin{equation}
H_\mathrm{tot} = \frac{P^2}{2 M}+ \frac{1}{2} M \Omega^2 Q^2 + 
\sum_{n=1}^N \left[\frac{p_n^2}{2 m}+\frac{m \omega_n^2}{2} (q_n-Q)^2 \right].
\label{htotal}
\end{equation}

Here the test particle, with generalized coordinates $(Q,P)$, is schematized as 
a harmonic oscillator of mass $M$ and angular frequency $\Omega$, while the 
generic $n^\mathrm{th}$ particle in the reservoir, with generalized coordinates $(q_n,p_n)$, 
is schematized as a harmonic oscillator of mass $m$ and angular frequency $\omega_n$. 
The systems are assumed to be one-dimensional, and the coupling
between the test particle and the particles in the reservoir has been chosen to be
translationally invariant, to avoid the appearance of renormalization terms as discussed 
in \cite{Hakim,Patriarca} (for more general couplings see also \cite{Hu}).

To illustrate the impact of the bath frequencies $\{\omega_n\}$, we
first consider two extreme situations for the range of their
distribution which are analytically approachable: infinite bandwidth 
and zero bandwidth. The first situation is the usual Langevin
approach, summarized here for completeness and to establish the notation
we will use in the following. The case of a heat bath with zero
bandwidth, i.e. all oscillators degenerate in frequency, is then 
studied, and we find that it produces a scaling law which can be 
approximated in the more realistic setting of a finite bandwidth.

\subsection{Heat bath with infinite bandwidth: Langevin scenario}

We briefly recall the major features of the Langevin approach and how
it may be derived through the analysis of an ensemble of harmonic 
oscillators with the proper energy distribution.  
By writing the Hamilton equations from (\ref{htotal}) and solving 
for the particles of the heat bath, the test particle can be described 
by the generalized Langevin equation \cite{Mori,Kubo}:

\begin{equation}
M \ddot{Q}(t) + \int_{t_0}^{t} ds \Gamma(t-s) \dot{Q}(s)+M \Omega^2 Q(t)=\Pi(t)
\end{equation}

\noindent 
where the kernel of the dissipative term (in general non local in time) is:

\begin{equation}
\Gamma(t-s)= \sum_n m \omega_n^2 \cos[\omega_n (t-s)]
\end{equation}

\noindent
and the fluctuation force term is:

\begin{equation}
\Pi(t) = \sum_n m \omega_n^2 \{
(q_n(t_0)-Q(t_0))\cos[\omega_n(t-t_0)]+  
\frac{p_n(t_0)}{m \omega_n} \sin[\omega_n(t-t_0)] \} 
\end{equation}

\begin{figure}[t]
\centering
\includegraphics[clip,width=0.80\columnwidth]{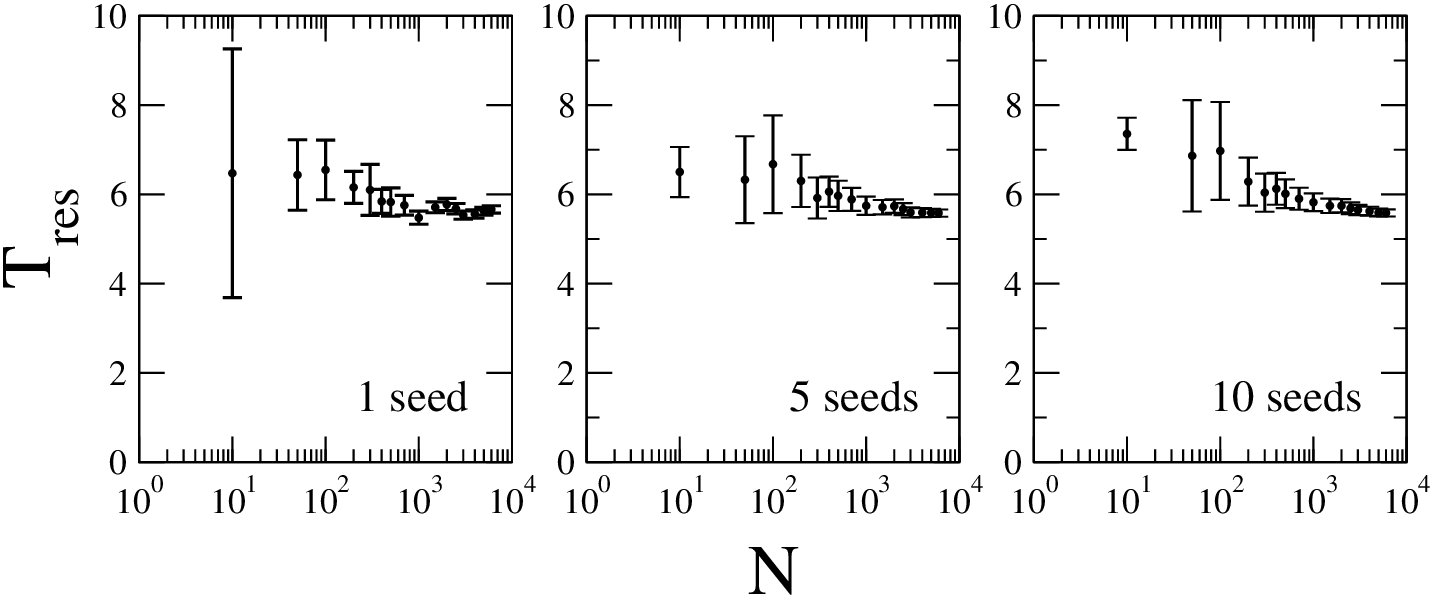}
\caption{Temperature of the heat bath versus the number of oscillators
for three runs differing by the number of seeds. 
The bath temperature, determined by the fitting of each
energy distribution with a Boltzmann  curve, is the weighted average of the
temperatures for each trial. The accuracy in the determination of the temperature
increases significantly by increasing the number of oscillators and
the number of seeds. In order to maintain reasonable times for the
numerical experiments, we have chosen to work with 400 oscillators and 5-7 seeds
in most of the single bath simulations.}
\end{figure}

In the limit of infinite harmonic oscillators in the bath extending
over the continuum frequency range $\omega_n \in [0, +\infty)$, with 
density of states $d N/d \omega \propto 1/\omega^2$, and if the 
initial conditions for the bath particles $(q_n(t_0), p_n(t_0))$ 
are chosen to realize a Boltzmann energy distribution, the test 
particle will be described by a Langevin equation (for details see \cite{Presilla})

\begin{equation}
dQ(t) = P(t) dt / M ; dP(t) = -[\gamma P(t)+M \Omega^2 Q(t)] dt + \sqrt{2 m \gamma k_B T} dw(t) 
\label{Langevin}
\end{equation}

\noindent
where $\gamma= \pi m \omega^2/2M \times dN/d \omega$. The stationary 
solution of Eqn. \ref{Langevin} corresponds to the thermalization of 
the test particle energy to the Boltzmann energy distribution of the bath.  

\subsection{Degenerate heat bath: zero bandwidth}

In order to understand the importance of having the oscillators of the bath distributed  
in a large bandwidth we also discuss the extreme case of a test particle interacting with 
a finite number of oscillators which are degenerate in frequency ($\omega_n =\omega_R$). 
In this case, if the initial conditions for the bath oscillators are distributed 
symmetrically in the phase space for each realization of the bath 
(such that $\sum_n q_n(t_0) \simeq 0$, $\sum_n p_n(t_0) \simeq 0$), 
the expressions for the dissipative kernel and the fluctuation force
become, apart from stochastic corrections due to the partial
cancellation in the initial conditions:

\begin{equation}
\Gamma(t-s)= N \omega_R^2 \cos[\omega_R (t-s)]
\end{equation}

\begin{equation} 
\Pi(t)= - N m \omega_R^2 Q(t_0) \cos[\omega_R(t-t_0)] 
\end{equation}

The Langevin equation in this case is then written as:

\begin{equation}
\ddot Q(t)- \xi \omega_R^3 \int_{t_0}^t Q(s) \sin[\omega_R(t-s)] ds+(1+\xi) \omega_R^2 Q(t)=0,
\end{equation}
where $\xi=Nm/M$. The corresponding dynamics resembles that of multiple oscillators 
experiencing beating phenomena, with energy periodically transferred between 
the test particle and the heat bath as $E_\mathrm{tp}(t)=E_0 \sin
\omega_R t$. The corresponding energy probability density $P(E)$ for
the test particle having energy between $E$ and $E+dE$ is therefore 
$P(E) \propto (dE/dt)^{-1} \propto [E(E_0-E)]^{-1/2}$.  
The absence of a bandwidth for the oscillators schematizing the heat
bath is responsible for the lack of thermalization, as the test
particle performs a deterministic, periodic motion, and in such a
situation no thermalization can be expected. Nevertheless, this case 
is relevant for understanding the dependence of the thermalization
upon the parameter $\xi$, as we will discuss in the
next section. 

\begin{figure}[t]
\centering
\includegraphics[clip,width=0.60\columnwidth]{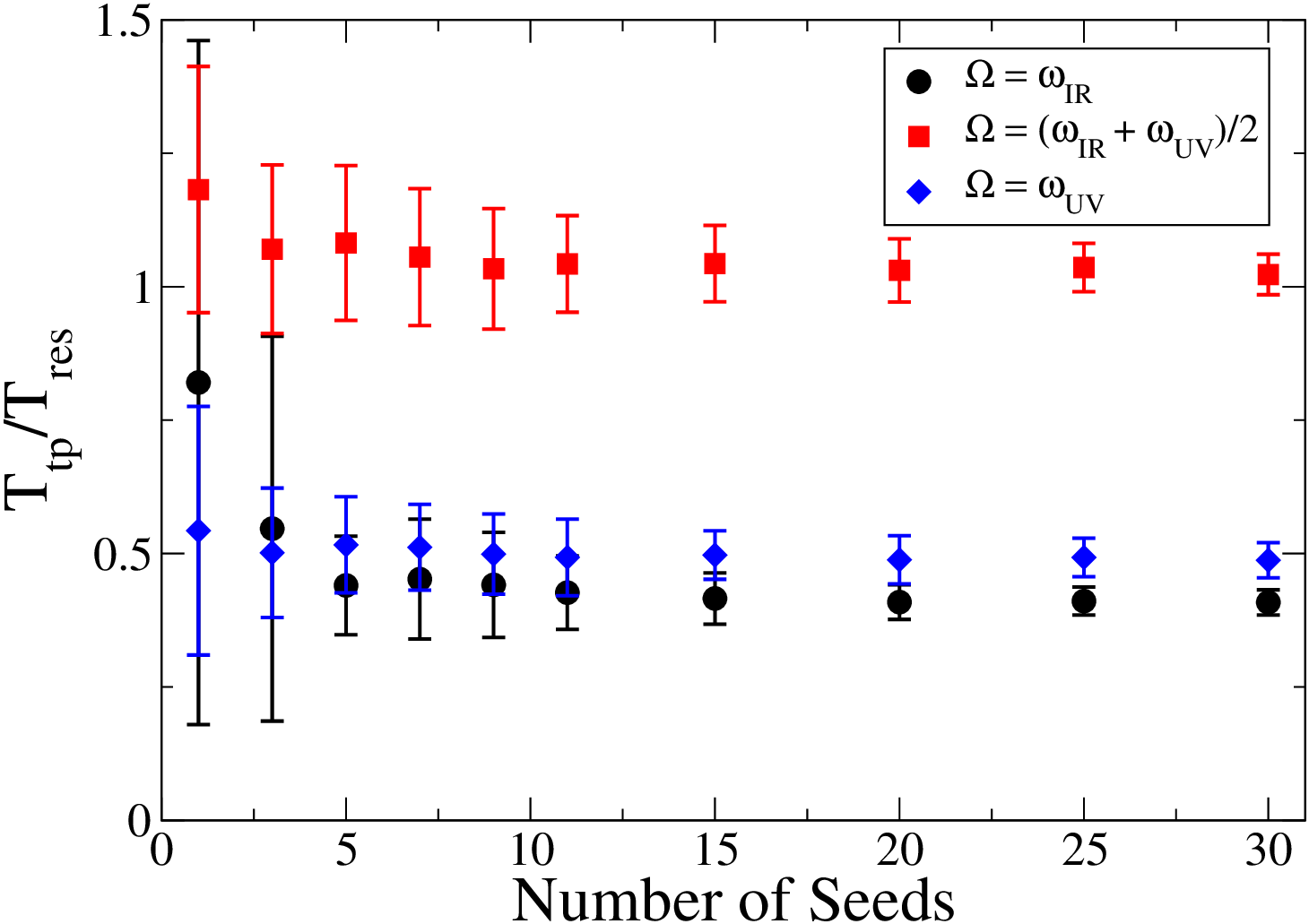}
\caption{(Color online) Temperature of the test particle in units of the temperature of the heat 
bath versus the number of trials, each one characterized by a different seed. The 
test particle is embedded in a bath made by 400 oscillators,
and three situations are reported with different frequency 
of the test particle.}
\end{figure}

\section{Single heat bath: numerical results}
The above situations are highly idealized, since they both correspond
to an \textit{infinite} number of oscillators in the environment, either
spread out over all possible frequencies or concentrated at a single frequency. 
As we have mentioned in the introduction there is a strong justification
for relaxing these assumptions and considering the generic case of a finite environment
with a finite bandwidth for the frequencies of its particles. 

In a generic situation with the oscillators of the heat bath distributed
in a finite bandwidth, no explicit solutions will generally be available. 
However, we can conjecture that the behaviour will lie in between the
two extreme cases of the idealized Langevin heat bath and the degenerate heat bath. 
To corroborate this qualitative claim, we have used the method
described below to numerically investigate the deviations from
thermalization for a finite size reservoir characterized by a finite frequency spectrum 
$\omega_n \in [\omega_\mathrm{IR}, \omega_\mathrm{UV}]$, a finite number 
of oscillators, a finite mass ratio $m/M$ between the masses of the
reservoir particles and the test particle, and a given density of states.
  
Strictly speaking, for a heat bath made of a finite number of
oscillators coupled to a test particle, there will be no approach to
equilibrium, as any initial state for the combined system will recur
within a finite time \cite{Carcaterra1,Carcaterra2,Carcaterra3,Celik}. 
If the number of oscillators in the bath is sufficiently large,
however, the corresponding recurrence time will be very long. 
Provided one is interested in studying the dynamics on a timescale much
shorter than this rather long recurrence time, an effective approach
to equilibrium describable in terms of the usual thermodynamic
quantities is viable. In this spirit, we focus our attention on the mutual
relationship between the parameters of the heat bath and the test
particle, to find the conditions under which the latter thermalizes
with the former.

\subsection{Numerical approach to a finite size heat bath} 

Our numerical approach consists of choosing values for the parameters
which describe the test particle and the bath, solving the motion
equations for the system, sampling the test particle's energy 
over time, and analyzing its distribution. 
The energies $\{E_n\}$ of the oscillators in the bath are chosen to
follow a Boltzmann distribution, defining the bath temperature. 
The frequencies  $\{\omega_n\}$ are chosen independently to follow a
given--typically uniform--density of states, bounded by both the
infrared and ultraviolet cutoffs. The $\{E_n\}$ and $\{\omega_n\}$
serve to constrain the initial conditions $\{(q_n(t_0), p_n(t_0))\}$,
which are then chosen to satisfy $E_n = \frac{p_n^2}{2m}
+\frac{1}{2}m\omega_n^2q_n^2$, each uniformly distributed in its
available phase space. All of these values are chosen using the MATLAB
pseudo-random number generator. In order to avoid bias among the
frequencies and energies of the bath oscillators, we run each
simulation multiple times, using different seeds for the generator,
and average the results.

\begin{figure}[t]
\centering
\includegraphics[clip,width=0.60\columnwidth]{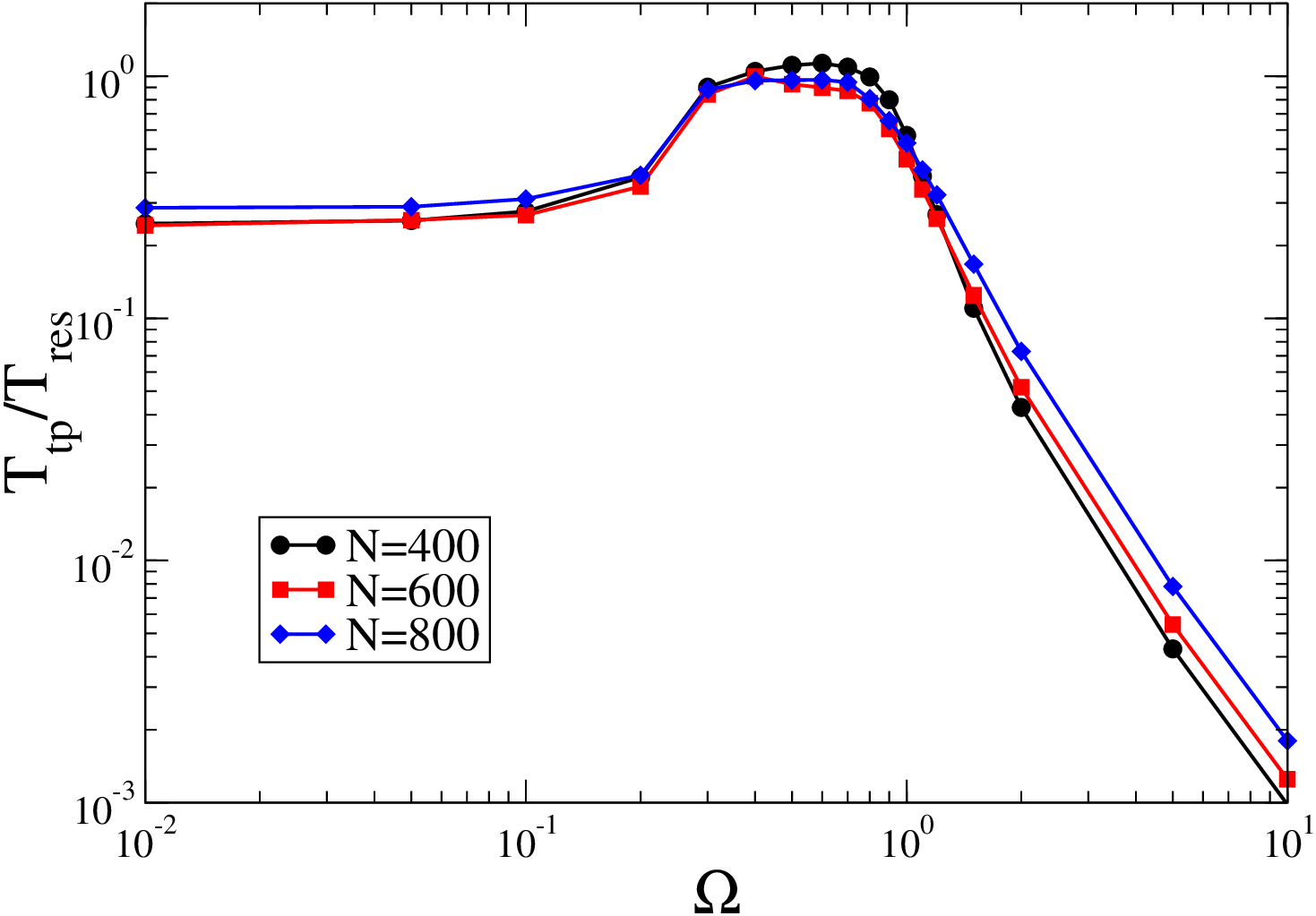}
\caption{(Color online) Thermalization plot with the temperature of the test particle
in units of the temperature of the heat bath versus its frequency, for three 
different numbers of oscillators in the bath, N=400, 600, and 800, and
15 seeds. The infrared cut-off frequency of the bath is $\omega_\mathrm{IR}$=0.2
and the ultraviolet cut-off frequency is $\omega_\mathrm{UV}$=1.}
\end{figure}

Having specified the quantities in the bath Hamiltonian, the original 
system of equations can be solved exactly -- at least to the level of 
precision at which a matrix can be diagonalized numerically -- using 
the following approach. The coordinates for each particle in the 
system are arranged into a vector $v$, with time derivative $\dot{v}$: 
\[ v(t) = \left( \begin{array}{c}
Q(t)\\
P(t)\\
q_1(t)\\
p_1(t)\\
\vdots\\
q_N(t)\\
p_N(t) \end{array} \right), \,\,\,\,\,\,
\dot{v}(t) = \left( \begin{array}{c}
\dot{Q}(t)\\
\dot{P}(t)\\
\dot{q_1}(t)\\
\dot{p_1}(t)\\
\vdots\\
\dot{q}_N(t)\\
\dot{p}_N(t) \end{array} \right) \]

\begin{figure}[t]
\centering
\includegraphics[clip,width=0.60\columnwidth]{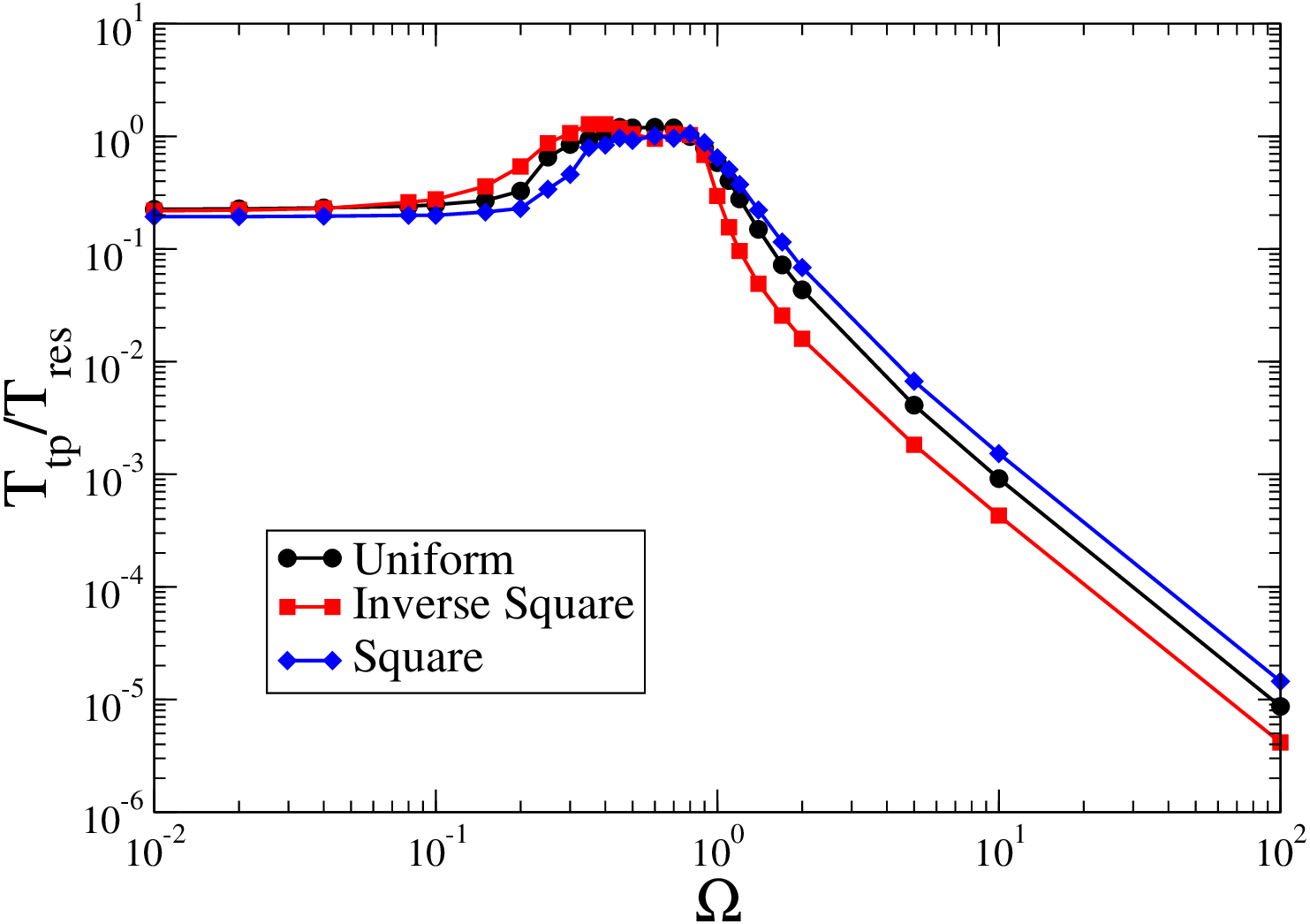}
\caption{(Color online) Thermalization curves for a test particle interacting with
  heat baths having frequencies in the range between 0.2 and 1,
  distributed uniformly, and with inverse square and square dependence
  on the frequency in the same bandwidth. The presence of more
  oscillators at lower or higher frequencies induces thermalization at
  lower or higher frequencies of the test particle, respectively.}
  \label{densitiesofstate}
\end{figure}

\begin{figure}[t]
\centering
\includegraphics[clip,width=0.60\columnwidth]{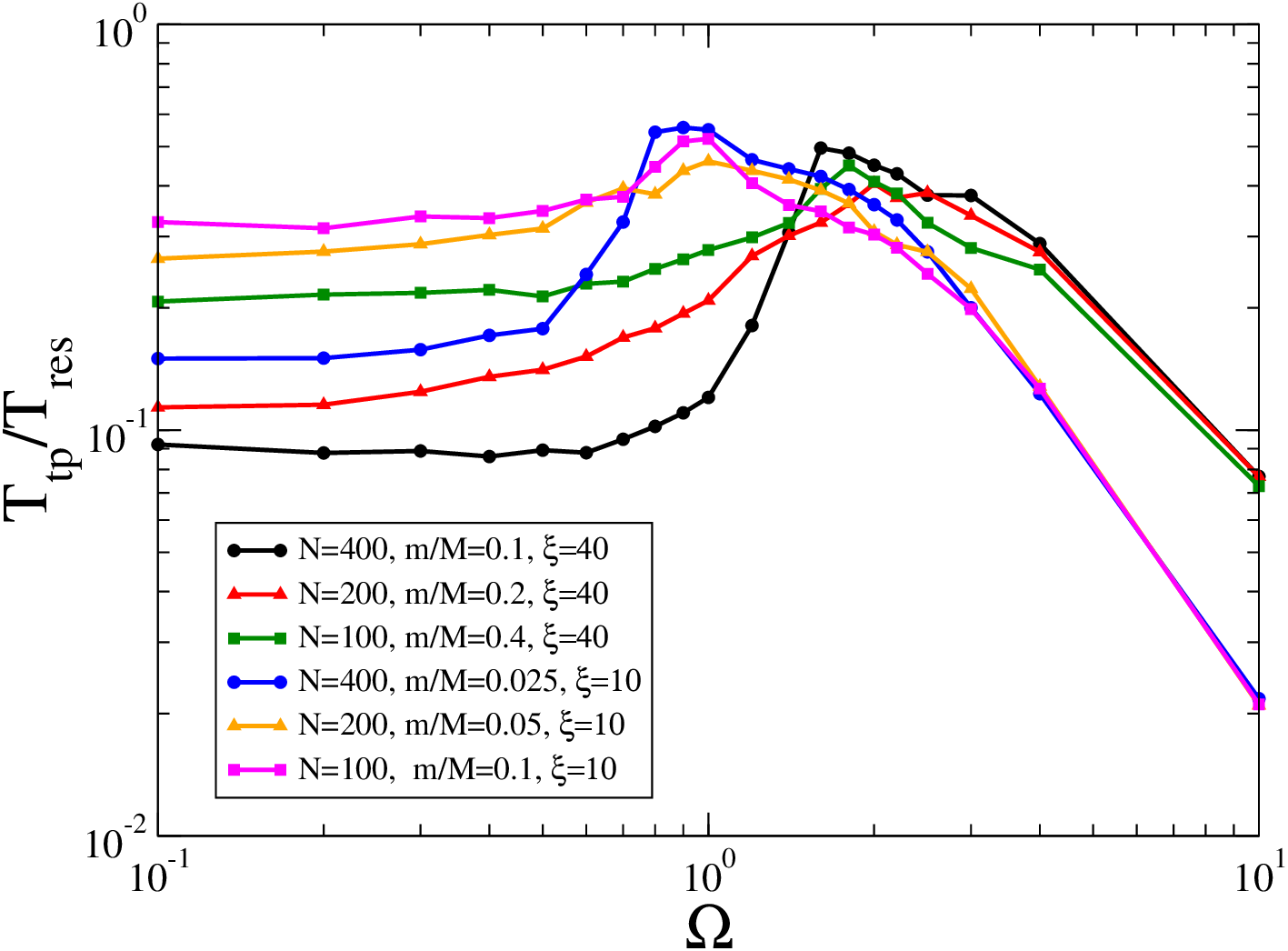}
\caption{(Color online) Approximated scaling properties of the thermalization peak
upon the parameter $\xi=Nm/M$. Shown is the ratio between the
temperature of the test particle and the heat bath temperature versus
the proper frequency of the test particle near the optimal
thermalization region for two different values of $\xi$. The 
three curves within each value of $\xi$ differs by the number of 
particles and the mass ratio, respectively $N$=100, 200, 400 
and $m/M$=0.1, 0.05, 0.025 for the case of $\xi$=10, and
$N$=100, 200, 400 and $m/M$=0.4, 0.2, 0.1 for the case of $\xi$=40.
Notice that for the same $\xi$ value the thermalization peaks occur at
approximately the same frequency. The case of a degenerate bath 
would predict a ratio between the frequencies at which the peaks
of thermalization occur as $\sqrt{41/11} \simeq 2$ in approximate
agreement with the observed pattern.}
\label{xidependence}
\end{figure}

Using this notation, the system of $2N + 2$ equations describing the 
system--the Hamilton equations from Eq. \ref{htotal}--can be rewritten in matrix form as 
\begin{equation} \label{mateq}
\dot{v} = Av 
\end{equation} 
where A is a square matrix of dimension $2N+2$ given by
\begin{equation}
 A = \left( \begin{array}{ccccccccc} \label{Amatrix}
0 & \frac{1}{M} & 0 & 0 & 0 & 0 &\cdots& 0 & 0\\
-M\Omega^2 - \sum_n m\omega_n^2& 0 & m\omega_1^2& 0 & m\omega_2^2& 0 &\cdots&m\omega_N^2&0\\
0&0&0&\frac{1}{m}&0&0&\cdots&0&0\\
m\omega_1^2&0&-m\omega_1^2&0&0&0&\cdots&0&0\\
0&0&0&0&0&\frac{1}{m}&\cdots&0&0\\
m\omega_2^2&0&0&0&-m\omega_2^2&0&\cdots&0&0\\
\vdots &\vdots&\vdots&\vdots&\vdots &\vdots& \ddots & \vdots & \vdots\\
0&0&0&0&0&0&\cdots&0&\frac{1}{m}\\
m\omega_N^2&0&0&0&0&0&\cdots&-m\omega_N^2&0\\
\end{array} \right). 
\end{equation}

This matrix can be written as $A = CDC^{-1}$, where $D$ is the diagonal 
matrix of eigenvalues, and $C$ is the matrix of corresponding eigenvectors. 
We can then rewrite the matrix equation (\ref{mateq}) as $ \dot{v} =
CDC^{-1} v $
and then $ C^{-1} \dot{v} = DC^{-1} v$.
If we let $ v^{'} = C^{-1}v$ and $\dot{v}^{'} = C^{-1}\dot{v}$ then we have 
a new matrix equation in normal coordinates $ \dot{v}^{'} = Dv^{'}$. 
This yields a set of independent, second order, homogenous differential equations
$\dot{v}^{'}_n = d_nv^{'}_n$ where the $\{d_n\}$ are the diagonal elements of $D$, 
the eigenvalues of $A$. Each of these equations has the simple solution 
$v^{'}_n = v^{'}_{0_n}e^{d_nt}$ where $v^{'}_{0_n}$ is the $n^{th}$
component of the vector of initial conditions in normal coordinates, 
obtained by applying $C^{-1}$ to the initial conditions $v_0$. 
We can write these solutions in a vector
\[ v^{'}(t)  = \left( \begin{array}{c}
v^{'}_{0_1}e^{d_1t}\\
v^{'}_{0_2}e^{d_2t}\\
  \vdots \\
v^{'}_{0_N}e^{d_Nt}
 \end{array} \right). \] 
The coordinate vector containing the position and momentum of the test
particle and bath particles at any time $t$ can then be obtained by
substituting the value of $t$ into $v^{'}(t)$ and multiplying by the
matrix $C$ to return to the original coordinates. 

Since $C$ has the same dimensions as $A$--
specifically, $(2N+2)\times(2N+2)$--this causes the process of
determining the state of the system at some time to run in time which
is $O(N^2)$ in the number of particles in the bath. However, it is
possible to avoid this issue of computational resources if we are
willing to sacrifice knowledge about the state of each bath particle at time $t$, and 
simply request the knowledge of the position and momentum $q$ and $p$
of the test particle. In this case we can simply multiply the vector 
$v^{'}(t)$ by the first row of $C$ to obtain the value of $q$, and 
by the second row of $C$ to obtain the value of $p$--then the
process runs instead in $O(N)$, which significantly speeds up the 
simulations for relatively large numbers of oscillators.

The thermal behavior of the test particle can be observed by sampling 
its position and momentum at different times and calculating its
energy, and then making a histogram of the measured energies. 
If the distribution of energies follows the Boltzmann law, we say the test 
particle has thermalized, and we can define a temperature for 
the test particle by looking at the slope of the distribution. 
This temperature can then be compared with the temperature 
of the original energy distribution of the bath particles. 
The slope of the distribution is evaluated in a separate 
function by taking the natural logarithm of the height of 
each bin, and then performing weighted least squares fitting,
with the weight of each data point being given by $N_i$, the 
original height of the bin. The statistical fluctuations in 
the energy are expected to have a Poisson distribution, 
which if centered around a value $x$ has a standard deviation 
of $\sigma = \sqrt{x}$, and thus the uncertainty in the number 
of elements $N_i$ in each bin is given by $\sigma_i = \sqrt{N_i}.$
Since the weights in a weighted least squares fit are given by $w_i =
\frac{1}{\sigma_i^2}$, the weights for fitting $\ln N$ are then $N_i$. 
The relatively large number of oscillators and the statistical
accuracy of the results do not require fitting with a canonical 
power law distribution as discussed in \cite{Plastino,Adib,Potiguar1,Prosper,Potiguar2}.

One issue in sampling the energy of the test particle is 
considering the time scales involved in the problem, and 
possible periodicity. Any sampling corresponding to
periodicities in the motion of the test particle may produce
misleading results. For this reason we use random step sizes 
in sampling the test particle energy, with the distribution 
of random step sizes centered around a chosen value, uniformly 
distributed between zero and twice that value, in such a way that the latter 
represents the average sampling time. 

The precision in the determination of the temperature of the heat bath
and the test particle are due to different factors. 
The precision of the bath temperature is limited by the number 
of oscillators it comprises, which is generally in the hundreds. 
Even generating multiple realizations of the bath energies and 
averaging over these to compensate for the quality of the distribution 
produced by the random number generator does not result in significant 
improvements here. This is shown in Fig. 1 where the heat bath 
temperature and its standard deviation are plotted versus the 
number of oscillators for three different numbers of realizations,
each of which represents a different seed for the random number generator.   
For the test particle temperature, on the other hand, the precision at leading
order is determined by the chosen number of random realizations. Averaging over a large 
number of trials lowers the influences from bad seeds when generating the 
energy and frequency distributions of the bath particles, as can be seen
in Fig. 2. However, too many trials will result in a very long simulation time. 
For those tests which require high precision we in general use 15
seeds, while for tests requiring less precision 5-7 seeds are used. 

\begin{figure}[t]
\centering
\includegraphics[clip,width=0.70\columnwidth]{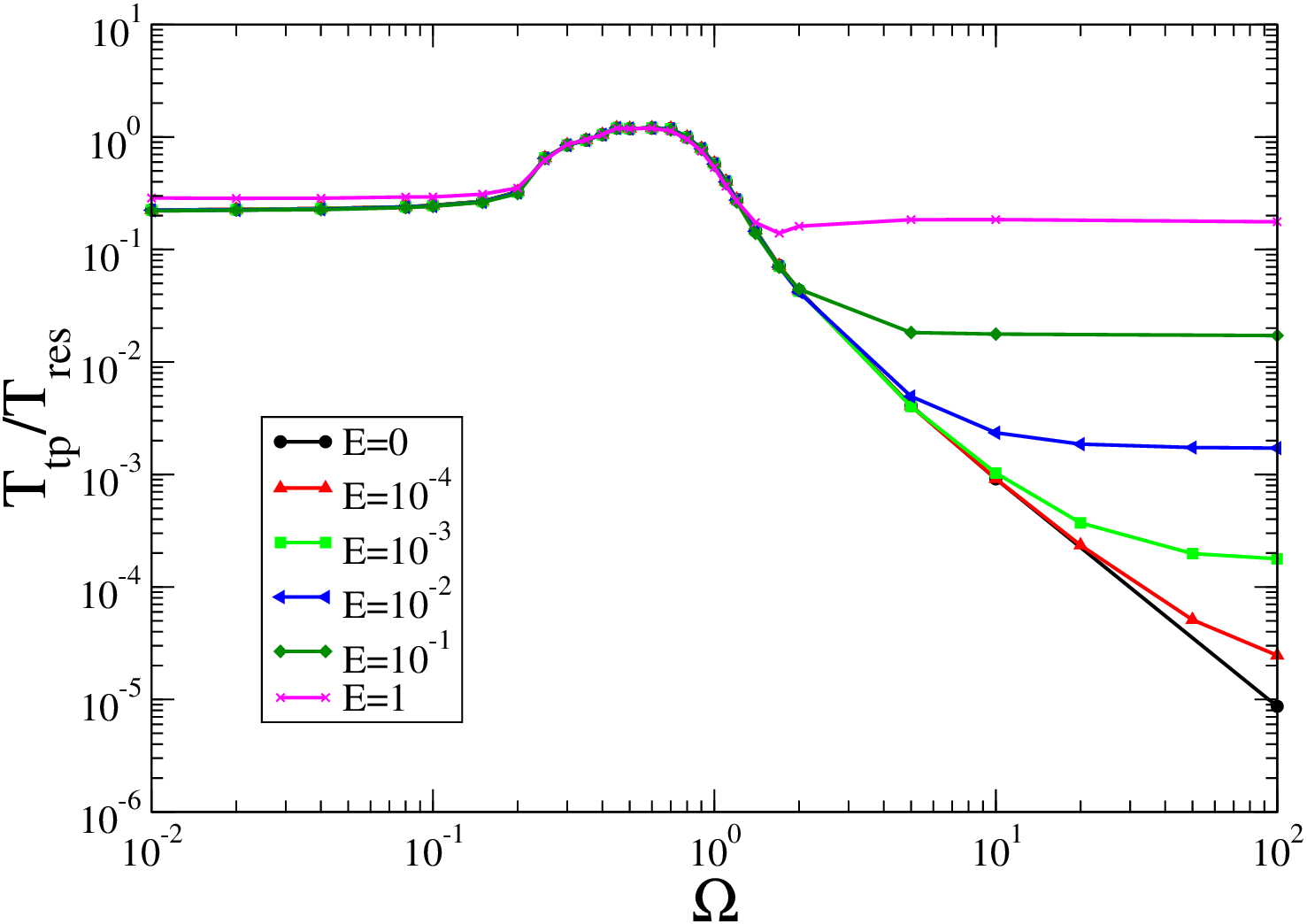}
\includegraphics[clip,width=0.70\columnwidth]{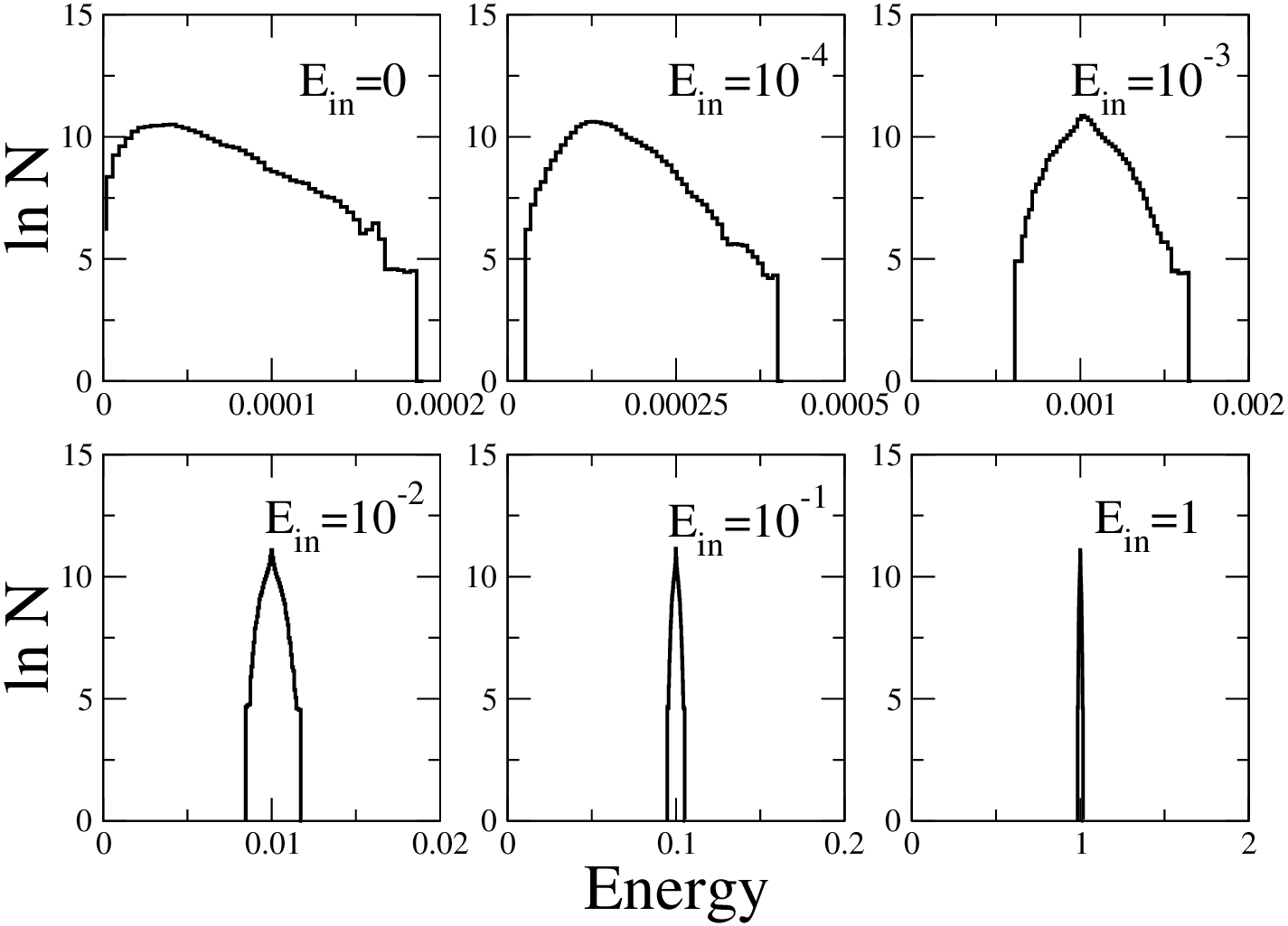}
\caption{(Color online) Top plot: thermalization curves for a test particle interacting with
heat baths having frequencies in the range between 0.2 and 1,
distributed uniformly, for different values of the test particle
initial energy. At high frequency the test particle is effectively
decoupled from the heat bath and therefore maintains its initial energy.
Bottom plot: energy distribution for the test particle with $\Omega=100$ 
versus its initial energy, with the bath temperature being $T$=5.5 in 
arbitrary units. The distribution appears Boltzmann at low 
energies, but becomes Gaussian at high energies, with its peak 
around the initial energy.}
\end{figure}

\section{Conditions for thermalization of the test particle to a heat bath}

Examples of partial and complete thermalization of the test 
particle to the reservoir are shown in Fig. 3 for different numbers
of oscillators in the heat bath (see \cite{Taylor} for examples of the test particle energy 
distribution). We find three regimes: at low angular frequencies 
$\Omega << \omega_\mathrm{IR}$, the test particle
approaches an equilibrium state corresponding to an average energy 
lower than the reservoir temperature, its spectral dependence is flat, 
and the energy distribution is significantly different from the
Boltzmann one, as discussed in \cite{Taylor}. 
In the intermediate regime  $\omega_\mathrm{IR} \leq \Omega \leq  
\omega_\mathrm{UV}$, the test particle generally approaches complete thermalization. 
In the region where $\Omega >> \omega_\mathrm{UV}$, the test particle 
thermalizes at a temperature significantly lower than the one of the 
reservoir, with a strong dependence upon the frequency. 
We have tested the dependence of the thermalization plot on the choice
of the density of states of the heat bath, as show in Fig. 4. 
Keeping all the remaining parameters constant, the density of states 
is chosen to be proportional to $1/\omega^2$, independent of $\omega$,
or proportional to $\omega^2$ in the three plots, in all cases within 
the same spectral interval (and zero otherwise). 
It is apparent that for the global thermalization features it makes 
little difference whether the frequencies follow a uniform, square, 
or inverse square distribution; the only difference is visible at 
the edges of the peak thermalization region, where the presence of 
more oscillators at lower or higher frequencies induces thermalization 
at slightly lower or higher frequencies of the test particle,
respectively. Considering the similarity among the various choices for
the density of states, we have used the simplest uniform distributions 
in all subsequent simulations. 

The thermalization region can be understood by thinking of the case 
of a degenerate heat bath: if the test particle is in resonance with 
the bath oscillators, it will trade energy with them in an efficient 
way, and otherwise the energy exchange will be limited both in 
amplitude and speed. That the test particle thermalizes generally 
near the bandwidth of the heat bath is thus to be expected, although 
in some cases the effective frequency of the test particle in contact with
the bath is sufficiently different than its natural frequency $\Omega$
that the region of peak thermalization is shifted. Since the effective
frequency under the effect of the heat bath is renormalized as 
$\tilde{\Omega}= (1+\xi)^{1/2} \Omega$, if $\xi$ is significantly 
larger than zero we expect the resonance to occur at angular frequencies 
higher than the intrinsic angular frequency of the test particle, 
and this should only depend on $\xi$. 
The shifts of the thermalization regions seen in Fig. 5 are in
agreement with this scaling hypothesis. This has as a further
implication a non monotonic dependence of the thermalization 
upon the mass ratio $m/M$, which cannot be trivially addressed 
in a perturbative approach \cite{Plyukhin}. 
As already discussed in \cite{Taylor}, the thermalization efficiency of the test
particle at a given oscillation frequency is initially directly
proportional to the mass ratio $m/M$ determining the coupling strength 
of the bath-test particle interaction, and then reaches an optimal
value before the $\xi$-dependent renormalized frequency plays a role.

Another phenomenon worthy of note occurs in the high frequency
region, where the thermalization curve is dependent upon the 
initial conditions of the test particle, as it is evident from
the thermalization plots in the top plot of Fig. 6. While at lower 
frequencies the particle thermalizes to the same temperature regardless
of its initial energy, at higher frequencies it is effectively
decoupled from the lower frequency heat bath. These energy distributions 
are reported in the bottom plots of Fig. 6. 
This non-Markovian behavior can be understood as the 
classical counterpart of the dynamical decoupling techniques developed  
to protect a quantum system from environmentally induced decoherence 
\cite{Viola}, in which a system driven at a frequency much higher than
the ultraviolet cut off of the bath is effectively decoupled. 
In this case the test particle oscillation frequency 
naturally provides a dynamical self-decoupling mechanism. 

\begin{figure}[t]
\centering
\includegraphics[clip,width=0.60\columnwidth]{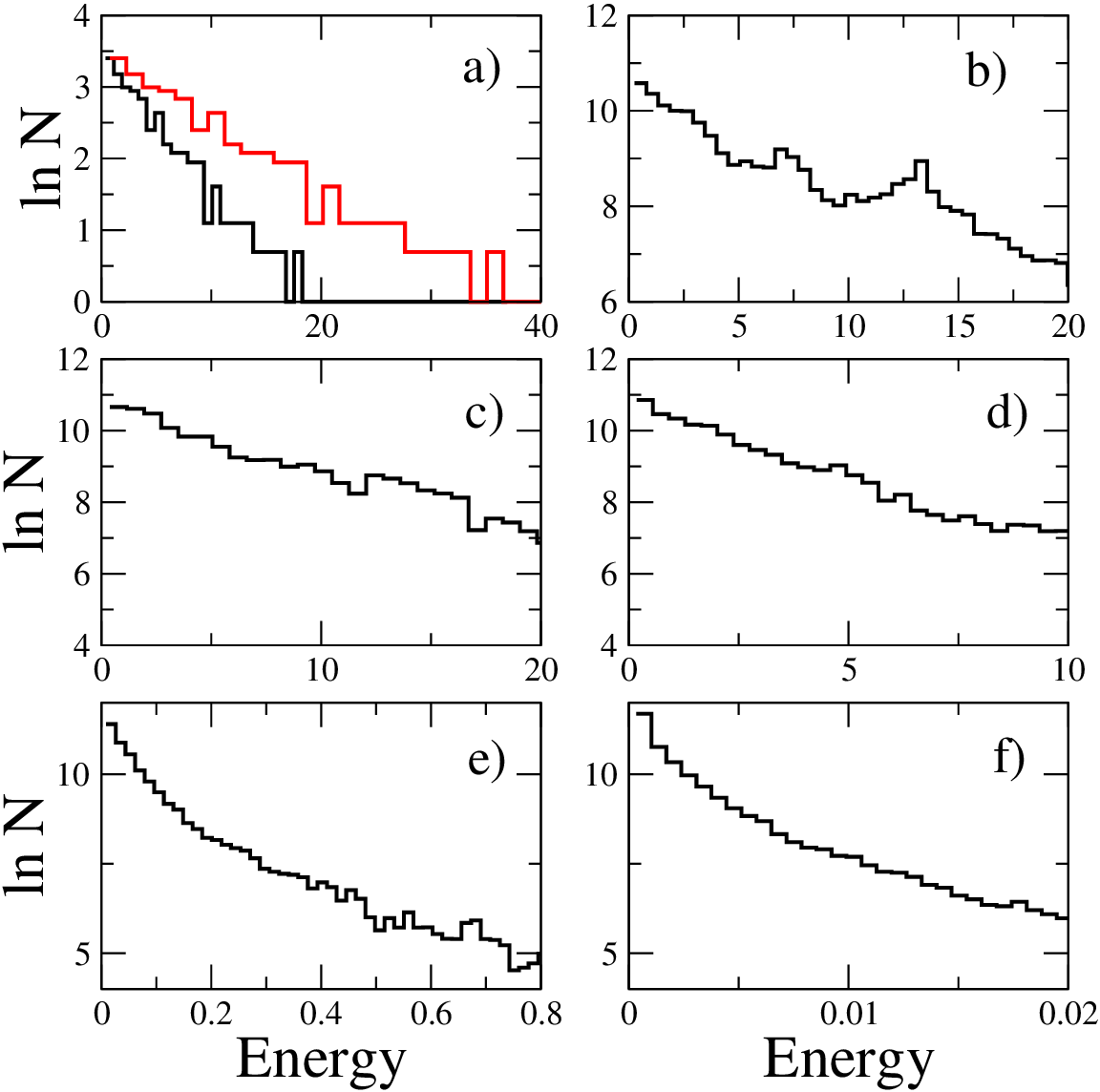}
\caption{(Color online) Energy distribution in the case of a test particle
intermittently interacting with two baths at different temperature. 
In the top left plot the energy distribution for the two baths at temperatures 5 and 
10 respectively is depicted, while in the other plots the energy distribution for 
the test particle for different values of its frequency is shown (from
b to f, $\Omega$ = 0.01, 0.3, 0.6, 2, 10). In all the simulations we
have used 200 oscillators in each heat bath, with a mass ratio between 
the bath particle and the test particle $m/M=10^{-3}$, and a switching
time $\Delta T=1$ in units of time steps.}
\end{figure}

\section{Dynamics in the presence of two thermostats} 

Here we discuss the case of a test particle connected to two heat
baths with finite resources. In order to avoid simultaneous contact 
between the two baths through the test particle, which will eventually
lead to their complete thermalization, we consider an intermittent 
interaction between the test particle and the baths. In this way, as 
also tested at the end of each simulation, the two baths still 
preserve their original temperature and energy distribution for 
a weak bath-test particle coupling. 

\begin{figure}[t] 
\centering
\includegraphics[clip,width=0.60\columnwidth]{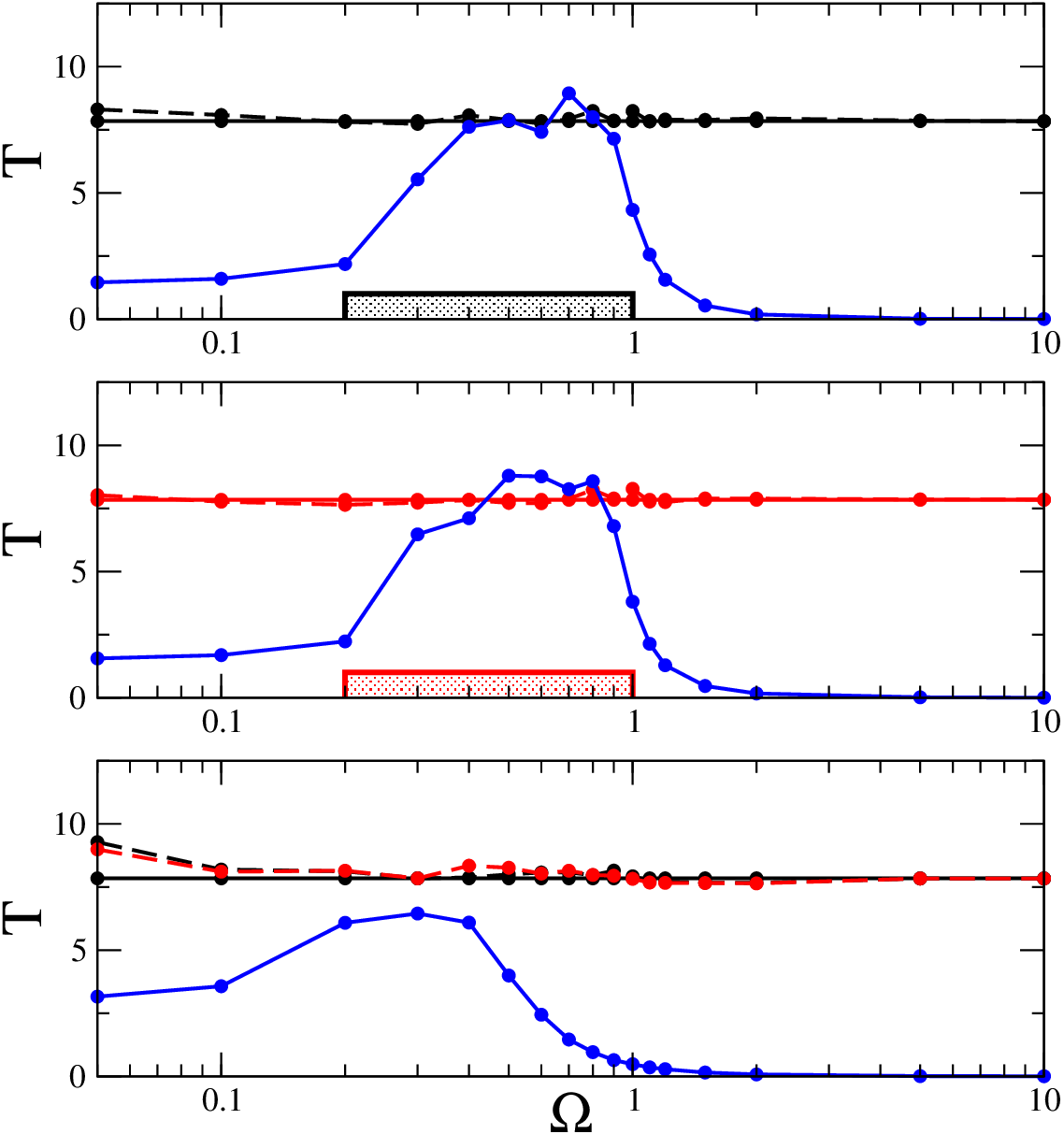}
\caption{(Color online) Thermalization plots for the case of two heat baths in interaction with 
a test particle versus its proper frequency. In the top plot the case
of the test particle only interacting with a heat bath at temperature
7.5 (in arbitrary units) is reported (blue curve), together with the
initial and final temperature of the heat bath acting on it (black
curves, continuous and dashed, respectively). In the middle plot the 
test particle is  interacting with another heat bath at the same initial
temperature of 7.5 and the same density of state as the lower
temperature bath. In the bottom plot the effect of the two baths
is shown when they intermittently interact with the test particle.
Notice that optimal thermalization occurs at frequencies lower than in
the case of each bath acting independently, leading to an effective
temperature lower than the common one of the two heat baths.}
\end{figure}

The total Hamiltonian for the system made of the two baths and the 
test particle is given by
\begin{equation}
H_\mathrm{tot} = \frac{P^2}{2M} + \frac{1}{2}M\Omega^2Q^2 + 
\sum_{n = 1}^{N1}\left[\frac{p_{1_n}^2}{2m} + \frac{1}{2}m\omega_{1_n}^2[q_{1_n}-\alpha_1(t)Q]^2\right] + 
\sum_{n = 1}^{N2}\left[\frac{p_{2_n}^2}{2m} + \frac{1}{2}m\omega_{2_n}^2[q_{2_n}-\alpha_2(t)Q]^2\right].
\end{equation}
\noindent
where $\alpha_1(t)=1$ if $2k \Delta T < t <(2k+1) \Delta T$ and
$\alpha_1(t)=0$ if $(2k+1) \Delta T < t < (2k+2) \Delta T$, and
$\alpha_2(t)=1-\alpha_1(t)$ (with $k$ an integer number), and
$\Delta T$ is the time the test particle is in contact with each bath.
By expanding the $(q_n -Q)^2$ terms, as in the one bath case, 
the Hamiltonian can be rewritten as

\begin{figure}[t] 
\centering
\includegraphics[clip,width=0.60\columnwidth]{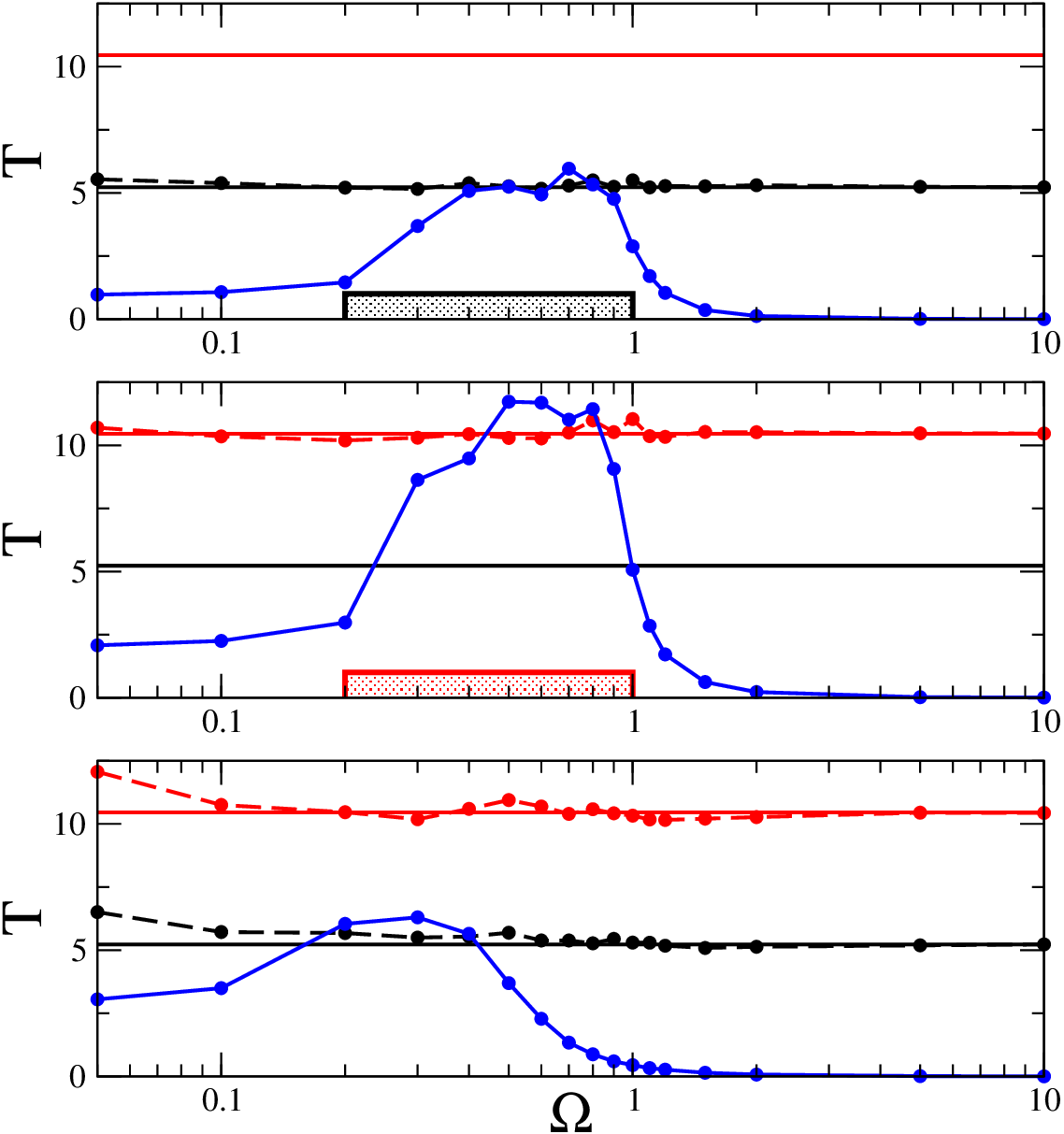}
\caption{(Color online) Thermalization plots for two heat baths in interaction with 
a test particle versus its proper frequency. In the top plot the case
of the test particle only interacting with a heat bath at temperature
5 (in arbitrary units) is reported (blue curve), together with the
initial and final temperature of the heat bath acting on it (black
curves, continuous and dashed, respectively). In the middle plot the 
test particle is instead interacting with a heat bath at the initial
temperature of 10, having the same density of state as the lower
temperature bath. In the bottom plot the effect of the two baths
is shown when they intermittently interact with the test particle.
Notice that optimal thermalization occurs at frequencies lower than in
the case of each bath acting independently.}
\end{figure}

\begin{eqnarray} \label{Htwobath}
H_\mathrm{tot} &=&  \frac{P^2}{2M} + \frac{1}{2}M\left[\Omega^2 + 
\alpha_1(t)^2 \sum_{n = 1}^{N1}\frac{m}{M}\omega_{1_n}^2 + 
\alpha_2(t)^2 \sum_{n = 1}^{N2}\frac{m}{M}\omega_{2_n}^2\right]Q^2 \nonumber \\
\, &+& \sum_{n = 1}^{N1}\left[\frac{p_{1_n}^2}{2m} + 
\frac{1}{2}m\omega_{1_n}^2q_{1_n}^2\right] \nonumber 
+\sum_{n = 1}^{N2}\left[\frac{p_{2_n}^2}{2m} + 
\frac{1}{2}m\omega_{2_n}^2q_{2_n}^2\right] \nonumber \\
\, &-& [\sum_{n = 1}^{N1}\alpha_1(t)m\omega_{1_n}^2q_{1_n} + 
\sum_{n = 1}^{N2}\alpha_2(t)m\omega_{2_n}^2q_{2_n}]Q
\end{eqnarray}

The second, third, and fourth terms in (\ref{Htwobath}) correspond
to the first bath independent of the test particle, the second bath 
independent of the test particle, and the interaction of the two baths 
with the test particle, respectively.

\begin{figure}[t] 
\centering
\includegraphics[clip,width=0.60\columnwidth]{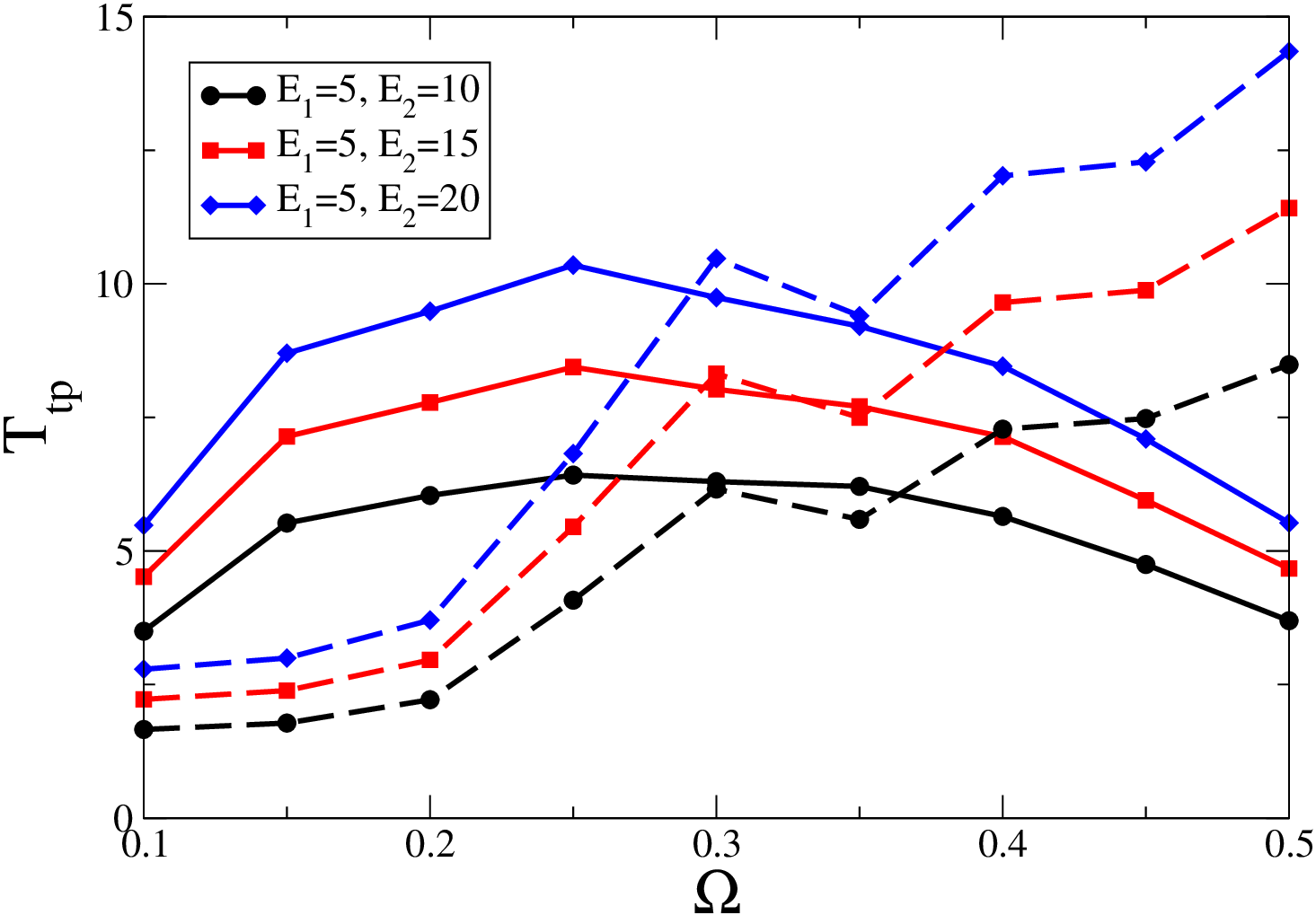}
\caption{(Color online) Thermalization plots for two heat baths in interaction with 
a test particle versus its frequency (solid lines), and comparison
with the arithmetic average of the two temperatures when the test 
particle interacts with the two baths separately (dashed lines). 
The colder bath is kept at temperature 5 (in arbitrary units), and the 
other bath is at temperature 10, 15 and 20 respectively for black, red and blue
lines, with a uniform density of states for the oscillators of the two
baths between 0.2 $\Omega$ and $\Omega$, and zero elsewhere. 
Notice that the resulting effective temperature of the test particle
is close to the arithmetic average of the temperatures of the two
baths only in a narrow region around 0.3-0.35 $\Omega$.}
\end{figure}

The equations of motion can be derived from the Hamiltonian, and 
the MATLAB code for the case of two baths operates in the same way 
as the one bath case.  The $\{q_{2_n}\}$ and $\{p_{2_n}\}$ for the 
second bath are simply appended to the vector of coordinates for 
the system $v$, so that $v$ and $\dot{v}$ have the form:

\[ v(t) = \left( \begin{array}{c}
\label{2bathmotion}
Q(t)\\
P(t)\\
q_{1_1}(t)\\
p_{1_1}(t)\\
\vdots\\
q_{1_{N_1}}(t)\\
p_{1_{N_1}}(t)\\
q_{2_1}(t)\\
p_{2_1}(t)\\
\vdots\\
q_{2_{N_2}}(t)\\
p_{2_{N_2}}(t) \end{array} \right), \,\,\,\,\,\,
\dot{v}(t) = \left( \begin{array}{c}
\dot{Q}(t)\\
\dot{P}(t)\\
\dot{q}_{1_1}(t)\\
\dot{p}_{1_1}(t)\\
\vdots\\
\dot{q}_{1_{N_1}}(t)\\
\dot{p}_{1_{N_1}}(t)\\
\dot{q}_{2_1}(t)\\
\dot{p}_{2_1}(t)\\
\vdots\\
\dot{q}_{2_{N_2}}(t)\\
\dot{p}_{2_{N_2}}(t) 
\end{array} \right) \]

\begin{figure}[t] 
\centering
\includegraphics[clip,width=0.60\columnwidth]{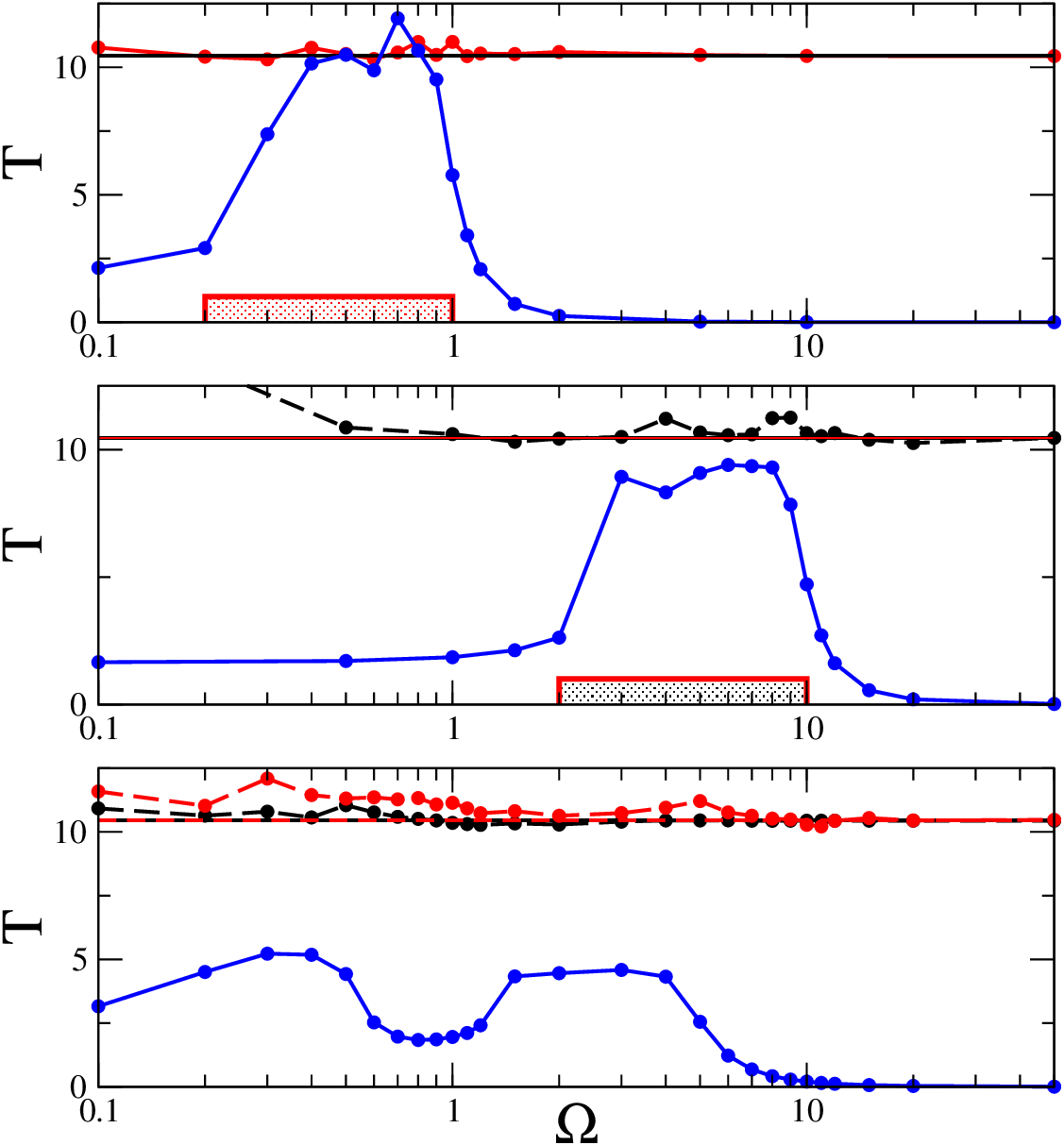}
\caption{(Color online) Thermalization plots for two heat baths in interaction with a
test particle versus its proper frequency. The two heat baths have
the same temperature 10 (in arbitrary units) and different,
non-overlapping, density of states (see top and center figures). 
When only one bath is acting, the test particles thermalizes to 
the corresponding bath if its frequency is in the same spectral 
region as the density of states of the bath.  In the bottom plot 
the effect of the two baths is shown when they intermittently 
interact with the test particle, with two different regions 
of partial thermalization.}
\end{figure}

\begin{figure}[t] 
\centering
\includegraphics[clip,width=0.49\columnwidth]{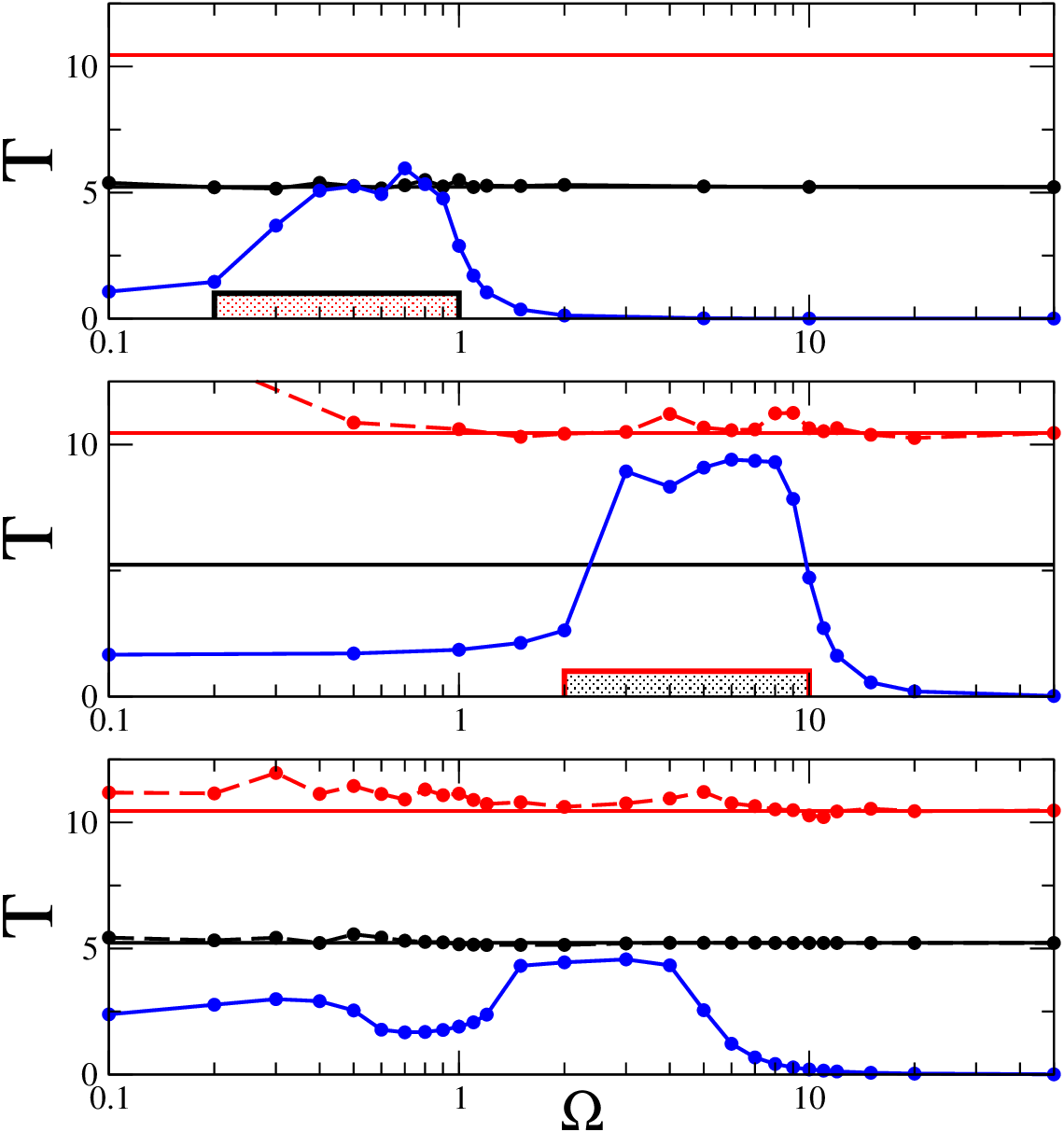}
\includegraphics[clip,width=0.49\columnwidth]{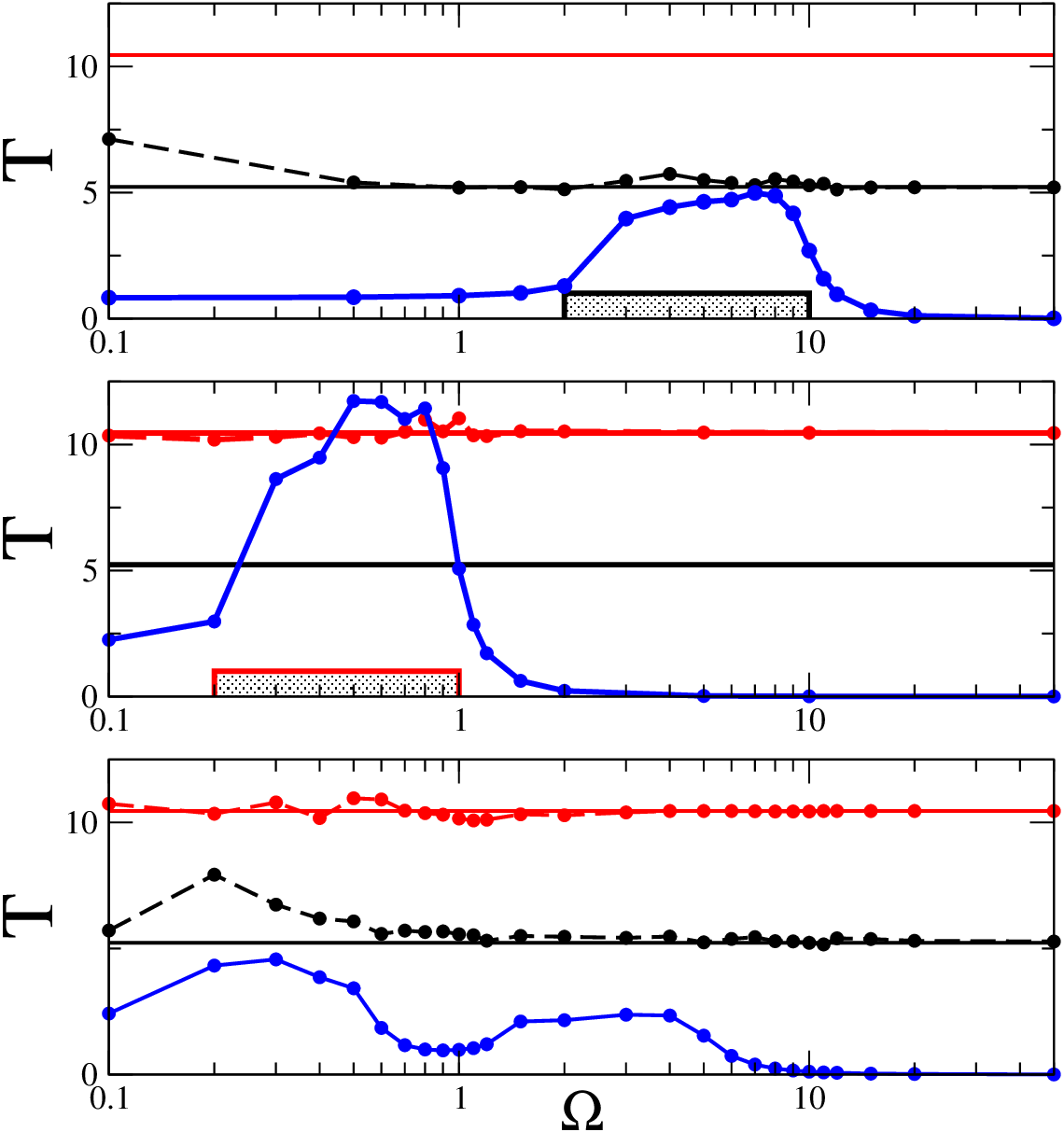}
\caption{(Color online) 
Thermalization plots for two heat baths in interaction with a
  test particle versus its proper frequency. The two heat baths 
have both different temperature and density of states, as visible in
  the top and center plots, with the colder bath (at T=5 in arbitrary
  units) having the bandwidth centered at lower frequency range (0.2 to 1) in the
  left plot, and centered at higher frequency range (2 to 10) in the right plot. 
  In the bottom plot the effect of the two baths
is shown when they intermittently interact with the test particle,
  which is frustrated both by the different temperatures and the
  different density of states.}
\end{figure}

The system of Hamilton equations describing the motion
of the system, now of size $2(N_1 + N_2) + 2$, can be replaced by 
the matrix equation $\dot{v} = Av$, where $A$ has the similar form 
as (\ref{Amatrix}), but now has dimension 
$\left[2(N_1 + N_2) + 2\right]\times\left[2(N_1 + N_2) + 2\right]$. 
Since in our test the two baths are not simultaneously in contact 
the test particle, two matrices are constructed: $A_1$ in which all
the terms of bath 2 are set to zero and $A_2$ in which all terms 
of bath 1 are instead set to zero. The parameters for each of the 
two baths are allowed to be specified independently. 
In principle, this also allows for an 
arbitrary partition of $N$ harmonic oscillators in $\ell$ different 
heat baths, each specified by different initial conditions and density of states. 
The matrix diagonalization method used in one-bath thermalization to solve the equations
will not work now since it assumes the matrix $A$ does not change over
time. In the two-bath case instead $A$ switches between $A_1$ and $A_2$, simulating the test
particle switching between the two baths.  
Therefore we use a fourth order Runge-Kutta method to solve the
equations, switching between $A_1$ and $A_2$ at each step ($\Delta
T=1$ in units of time steps of the Runge-Kutta code). This fast
switching is the closest scenario to the case of a continuous 
interaction of the test particle with the two baths, maximizing the 
frustration of the test particle in equilibrating to each. 
Simulations with switching times $\Delta T$ of 2, 10, and 20 time
steps have been also performed yielding indistinguishable energy 
distributions within the statistical error.  
In the case of slow switching, {\it i.e.} for $\Delta T$ much longer 
than the thermalization time, the dynamics is complicated by the 
fact that the test particle, once thermalized with one bath, will 
have an initial condition with finite energy which can create 
a situation similar to the one discussed around Fig. 6. Therefore 
thermalization may be actually affected by the energy acquired by 
the test particle prior to the interaction with the next bath. 
In the simplest case of two infinite bandwidth Langevin baths 
as the one discussed in Section II.A, a thermalization to one 
bath first, and then to the other bath, is expected instead, 
regardless of the test particle energy. 
In this case the energy distribution for the test particle in the 
long time limit is expected to be the sum of two Boltzmann 
distributions, {\it i.e.} 
\begin{equation}
P(E)=\frac{1}{2}[\beta_1^{-1} \exp(-\beta_1
  E)+\beta_2^{-1}\exp(-\beta_2 E)]
\label{noneq}
\end{equation}
with $\beta_i=(K_BT_i)^{-1}, i=1,2$ the inverse temperature.  
We have also verified using the one-bath case 
that for a sufficiently small step size the Runge-Kutta method 
reproduces the results obtained using numerical diagonalization.

Fig. 7 shows the energy distribution in the case of the test particle interacting with 
two baths at different temperatures. Similar to the one bath case shown in 
\cite{Taylor}, the energy distribution of the test particle reaches a Boltzmann distribution 
when the frequency of the test particle is within the finite frequency spectrum of 
the bath particles, while the distribution becomes progressively less Boltzmann when the 
frequency of the test particle moves outside the spectrum. Fig. 8 shows the 
effect of the two baths with nominal equal temperature and density of states. 
The usual single bath thermalization curves when each of the baths 
is acting on the test particle are depicted in the top and middle
plots, while the thermalization plot corresponding to their
intermittent action is shown in the bottom plot.
It is worth noticing that the optimal thermalization region of the test particle 
is shifted towards lower angular frequencies, and its temperature is slightly
lower than the common temperature of the baths. In close analogy to the one bath 
case, one can then change the temperature of the two heat baths while keeping their 
density of state equal to each other. As shown in Fig. 9, it is
confirmed that the test particle's thermalization region is shifted towards
significantly lower frequencies with respect to the thermalization curves for
each separate heat bath. Also, the test particle is observed to thermalize at a
temperature lower than the average of the temperatures of the two baths. 
The average of the temperatures of the two baths is the effective
temperature of the test particle expected for $E << K_BT_1,
K_BT_2$ in the adiabatic switching limit when Eq. \ref{noneq} holds,
since in that case the test particle effective temperature
$T_\mathrm{eff}$ defined via the relationship:
\begin{equation}
T_\mathrm{eff}\exp(-E/K_B T_\mathrm{eff}) =[T_1\exp(-E/K_B T_1) +T_2 \exp(-E/K_B T_2)]/2
\end{equation}
becomes simply $T_\mathrm{eff} \simeq (T_1+T_2)/2$ up to terms of
order $(E/K_BT_1)^2, (E/K_BT_2)^2$. 

If we examine the lower thermalization temperature of the 
test particle more carefully, it seems to be a direct result from the 
shift of the thermalization region. Since this shifted thermalization region is not the 
optimal region for single bath thermalization, the test particle does not thermalize
with each of the two baths effectively. If we focus on the new optimal thermalization region, 
we can see that, although the test particle does not reach the average temperature of
the two baths, it does get close to the average of two temperatures when the test particle 
thermalizes with the two baths separately in a narrow frequency range
as shown in Fig. 10. 

Both these features are also present in the complementary 
situation of two heat baths at the same temperature but with non 
overlapping density of states, as shown in Fig. 11. In this case, while 
the test particle thermalizes at the common temperature of the heat baths when they 
are acting independently, in the intermittent case the two thermalization regions
are redshifted and the test particle's temperature is nearly half of the heat baths'. 
One possible explanation for this half thermalized temperature is that
the test particle cannot absorb much energy from the bath when 
its frequency is outside the frequency band of the bath as it has been shown in the one
bath case. Therefore a low frequency test particle does not receive energy from
the particles in the high frequency bath, and thus it is as if the
test particle is interacting with one bath at temperature 10 and the 
other bath at temperature 0, and likewise for the high frequency test 
particle. In other words, what matters is a sort of effective spectral
temperature for the test particle, and if there are no oscillators
around its proper frequency the corresponding bath is effectively at zero
temperature. In Fig. 12 we show the combined effect of heat baths at
different temperatures and different density of states. In the left
plot, we display the case of the cold bath having oscillators at
lower frequency (and the hot bath with density of states shifted
towards higher frequencies), with the test particle showing the
expected thermalization to each bath at the expected temperature and
in the same spectral region. When both baths are acting, the test
particle displays similar features to the previous figures, {\it i.e.}
an effective temperature about half of the temperature of the bath 
operating in the same frequency region, and a redshift of the
thermalization spectrum. In the right plot, we  present the dual 
situation, with the cold bath having oscillators at higher frequency. 

\section{Conclusions}

We have studied finite resource reservoirs, showing that
thermalization of a test particle occurs only under specific 
conditions in both the case of continuous interaction with a single bath and 
intermittent interaction with two baths with different features. 
These results may have an impact on the investigations of equilibrium and 
nonequilibrium quantum statistical mechanics of nanostructures, as 
first discussed in \cite{POB} in the framework of quantum-limited 
electromechanical tunneling transducers. Since then, many groups 
have pushed the sensitivity of displacement transducers for 
micro and nanoresonators near the quantum limit \cite{Braginsky}.
Three areas of current research could benefit from our discussions.
In \cite{Schwab}, a phenomenon of cooling has been interpreted as 
due to quantum back-action of the read-out system. A more economical 
interpretation, without necessarily invoking the quantum features of
the measurement apparatus, is instead available by imagining the 
effective temperature of the nanoresonator as resulting from the competition 
between two effective heat baths at different temperatures. 
This is also in line with the phenomenon of cold damping introduced eight decades 
ago \cite{Ornstein}, and demonstrated for macroscopic resonators in 
\cite{Hirakawa}. From our perspective, the study of the thermalization 
of a particle in simultaneous interaction with two heat baths should 
shed light on this phenomenon already at the classical level, especially 
focusing on the energy distribution which, in the presence of finite 
resources, is not necessarily of Boltzmann nature, and it could help
to understand some intriguing features of the effective temperature 
found in the demonstration reported in \cite{Schwab}. 
A second class of experiments which may benefit from our discussion 
are the ones involving very high frequency nanoresonators, as 
pioneered in \cite{Mohanty}. In our framework the existence of
discrete jumps could be attributed to non-equilibrium features of 
the test particle, in this case a high-frequency mode of a
nanomechanical structure. Due to its geometry and size, 
the nanoresonator may have a density of states of the phonon bath 
with Debye frequency much less than in the corresponding bulk situation. 
This should be confirmed by both dedicated experiments on thermal
properties, such as specific heat and heat conductivity, and 
{\it ab initio} calculations of the density of states. 
Our conjecture is that for lower dimensionality structures the Debye
frequency and the larger amount of relevant states may lie below 
the GHz range, {\it i.e.} below the resonator frequency.
The effective temperature felt by the test particle could then be 
quite different from the one expected by measuring the external bath 
to which the nanoresonator is coupled. Non-equilibrium distributions 
are also expected in the case, not considered here, of test particles
schematized in the quantum realm, for which even with a proper 
choice of observables deviations from the Boltzmann energy
distribution should be observed \cite{Onofrioepl}. 
Finally, we believe that the careful design of some environments 
could help to extend quantum nondemolition measurements \cite{Caves,Bocko} in
nanomechanical devices. It is well known that quantum nondemolition 
measurements are suitable only for the case of harmonic oscillators. 
Their extension to free particles in the form of measurements on
carefully prepared contractive states \cite{Yuen,Ozawa} or Schroedinger cat
states \cite{VO} is quite problematic in practice. Furthermore, the 
extension of quantum nondemolition measurements to generic nonlinear
systems is also difficult, as it requires a specific knowledge of the 
initial state of the system \cite{MOP}. We believe that proper
engineering of the environment surrounding the nanoresonator, with 
blockade of some frequency range for the exchanged phonons \cite{Patton,Geller,Qu,Wilson},
corresponding in our model to a proper preparation of the density of
states of the bath, will provide a robust and general way to reach 
and surpass the standard quantum limit.  

\acknowledgments
S.T.S. acknowledges partial support from 
the Mellam Foundation at Dartmouth.

\end{document}